\documentclass[a4paper,fleqn,usenatbib]{mnras}

\usepackage{newtxtext,newtxmath}

\usepackage[T1]{fontenc}
\usepackage{ae,aecompl}

\usepackage{graphicx}	
\usepackage{amsmath}	
\usepackage{amssymb}	
\usepackage{subfig}
\usepackage{eucal}
\usepackage{threeparttable}
\usepackage{makecell}

\title[Selection Effects in BHXBs.]{Selection Effects on the Orbital Period Distribution of Low Mass Black Hole X-ray Binaries }

\author[Arur \& Maccarone]{
K. Arur,$^{1}$\thanks{E-mail:kavitha.arur@ttu.edu}
and T. J. Maccarone$^{1}$
\\
$^{1}$Department of Physics and Astronomy, Texas Tech University,Lubbock,TX,79409-1051,USA\\
}

\date{Accepted 2017 October 19. Received 2017 October 18; in original form 2017 June 13}

\pubyear{2017}

\begin{document}
\label{firstpage}
\pagerange{\pageref{firstpage}--\pageref{lastpage}}
\maketitle
\begin{abstract}
We investigate the lack of observed low mass black hole binary systems at short periods ( < 4 hours) by comparing the observed orbital period distribution of 17 confirmed low mass Black hole X-ray Binaries (BHXBs) with their implied period distribution after correcting for the effects of extinction of the optical counterpart, absorption of the X-ray outburst and the probability of detecting a source in outburst. We also draw samples from two simple orbital period distributions and compare the simulated and observed orbital period distributions. We predict that there are >200 and <3000 binaries in the Galaxy with periods between 2 and 3 hours, with an additional $\approx$600 objects between 3 and 10 hours. 
\end{abstract}




\section{Introduction}

Low Mass X-ray Binaries (LMXBs) are binary systems consisting of a black hole or a neutron star primary and a low mass main sequence or an evolved companion star that is \textless 1$M_{\odot}$ (see e.g \citealt{2006csxs.book..157M}). Mass transfer occurs from the secondary to the primary via Roche lobe overflow. All of the dynamically confirmed low mass Black Hole X-ray Binaries (BHXBs) that have been observed are transient sources, spending most of the time in quiescence and occasionally going into outburst, when the source increases in X-ray luminosity by several orders of magnitude. The origin of the outbursts of transients is explained, at least in broad brush strokes, by the disc instability model (see reviews by \citealt{2001NewAR..45..449L} and   \citealt{2014SSRv..183..477M}).  The lack of persistent low mass BHXBs may be, in part, due to selection effects. For example, the persistent source 4U 1957+11 is likely to be a black hole (\citealt{2002MNRAS.331...60W}, \citealt{2015ApJ...809....9G}), but dynamical confirmation has not yet been obtained because the accretion disc outshines the donor star in optical wavelengths; still there are few strong candidate black holes in low mass X-ray binaries that are persistently bright sources. \\

\begin{figure}
	\includegraphics[width=\columnwidth]{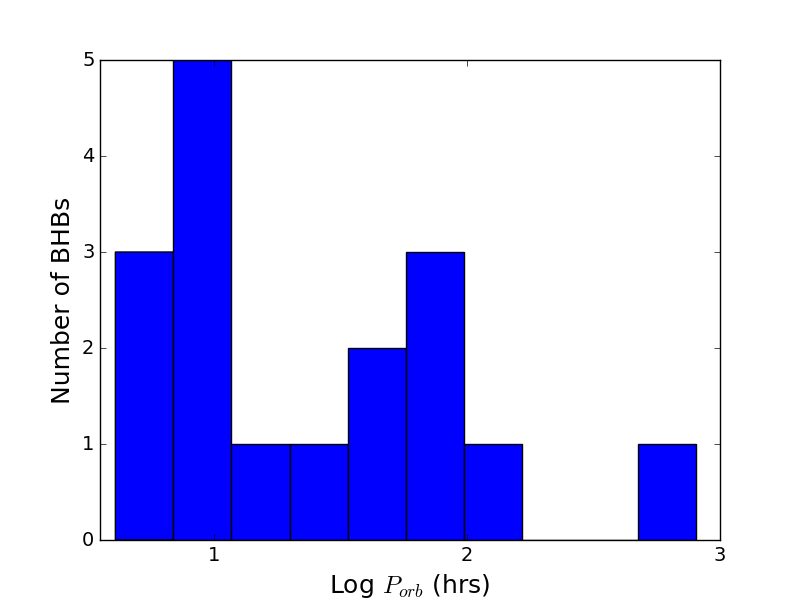}
    \caption{Histogram of the observed orbital period distribution of BHXBs. For a list of the properties of the binaries, see Table \ref{tab:bh_prop}.}
    \label{fig:Obs_dist}
\end{figure}

The orbital periods of these observed BHXBs vary from a few hours to several weeks (See Fig. \ref{fig:Obs_dist}). With the exception of GRS1915+105, all  BH systems have an orbital period shorter than a week. However, there appears to be a lack of BHXBs at the shorter periods (< 4 hours) relative to theoretical predictions (\citealt{1992ApJ...399..621R}, \citealt{1994ASPC...56..196R}, \citealt{1997A&amp;A...321..207P}, \citealt{2006A&amp;A...454..559Y}). \\

The peak X-ray luminosity of LMXBs during an outburst is directly proportional to the orbital period of the system (\citealt{1998MNRAS.301..382S}, \citealt{2004MNRAS.355..413P}, \citealt{2010ApJ...718..620W}). This is a result of the binaries with a longer orbital period having a larger disc radius, and a larger radius to which the accretion disc is irradiated by the central compact object. \\

When the mass accretion rates onto the primary are low and the local cooling timescale becomes longer than the accretion timescale, the accretion flow is radiatively inefficient. In this case an advection dominated accretion flow (ADAF) can occur (\citealt{1977ApJ...214..840I} , \citealt{1995ApJ...452..710N}) and much of the kinetic energy in the accretion flow is advected onto the compact object. In the case of a black hole primary, this energy is advected beyond the event horizon. In the case of a neutron star primary, however, the energy is radiated when the accretion flow impacts the stellar surface. This makes short period BHXBs intrinsically fainter than their neutron star counterparts \citep{2001ApJ...553L..47G}. \\

It was shown by \citet{2014MNRAS.437.3087K} that the lack of short period BH systems could be explained by a drop to a  radiatively inefficient state at short periods due to the lower disc mass in these systems. In this case, the BHXBs have a lower peak luminosity in outburst, shorter outburst durations and lower X-ray duty cycles that result in the system being much harder to detect. They also calculated the detection probabilities of BHXBs in the Galaxy as a function of their orbital periods to show that this switch to a radiatively inefficient state renders these short period systems undetectable at large distances. \\

In this paper, we investigate the impact of selection effects on the orbital period distribution of BHXBs by randomly distributing BHXBs across the Galaxy and calculating the fraction of sources that can be detected. We incorporate the switch to a radiatively inefficient state by assuming a gradual reduction in the radiative efficiency below a few percent of the Eddington luminosity as stated in \citet{2014MNRAS.437.3087K}. In addition to this switch, we also take into account the effects of absorption of the X-ray flux to model the likelihood of the detection of the X-ray outburst. \\

LMXBs tend to be concentrated towards the plane of the Galaxy, where the extinction levels are the highest. Radial velocity measurements obtained when the system is in quiescence are required to get dynamical mass estimates of the black hole candidate. In the case of short period binaries which have less luminous companion stars than do longer period binaries, these high levels of extinction can result in the optical counterpart being too dim for these measurements. Thus we have also included the probability of detecting the optical counterpart during quiescence in our calculations. \\
 
It is important to study and understand these effects because an accurate estimate of the population of BHXBs in the Galaxy can lead to more robust constraints on binary evolution models since the same physical process dominates the evolution of binaries irrespective of whether the primary is a white dwarf, neutron star or a black hole. In this paper,we study the implied period distribution of BHXBs and estimate the total number of BHXBs in the Galaxy. In Section 2 we present our sample of BHXBs and the implied period distribution after correcting for various selection effects. In Section 3 we compare the implied period distribution to the period distribution drawn from 2 sample initial orbital period distributions and accounting for the effects of magnetic braking. This is followed by a discussion of our results and our conclusions. 

\section{Data set and correction of selection effects}
\subsection{Galaxy model and data set}
\label{sec:galmod} 

The data set consists of the 15 confirmed BH LMXBs and 2 strong candidate sources that have since been dynamically confirmed listed in \citet{2006csxs.book..157M}. The sources and their properties are listed in Table~\ref{tab:bh_prop}. In order to determine what fraction of BHXBs are detected and dynamical mass measurements are obtained, 25000 locations were randomly selected in the Galaxy. The random selection was weighted by stellar density, making it more likely for a binary to be located in the denser parts of the Galaxy. Each of the 17 BHXBs listed in Table~\ref{tab:bh_prop} was then placed at each of the 25000 locations to test for detectability. SWIFT J1753.5-0127 has not been included in our data set due to controversy over whether it really has been dynamically confirmed \citep{2016MNRAS.463.1314S}. \\

\begin{table*}
\caption{Table of properties for the 17 BH LMXBs. }
\label{tab:bh_prop}
\begin{threeparttable}
\begin{tabular}{lccccl} 
		\hline
		Source Name & Alternate Name & $P_{orb}$(hr) & $m_V$ \tnote{a} & $A_V$  & Reference\\
		\hline
		GRO~J0422+32 & V518 Per & 5.1 & 22.4 & 0.74  & \citet{1995ApJ...455..614F}, \citet{2003ApJ...599.1254G}  \\
		A0620-00 & V616 Mon & 7.8 &  18.2 & 1.21  & \citet{1976Wu}, \citet{1977ApJ...217..181O}, \citet{1986ApJ...308..110M} \\
		GRS~1009-45 & MM Vel & 6.8 & 21.7 & 0.62 & \citet{1999PASP..111..969F}, \citet{Masetti}\\
		XTE~J1118+480 & KV UMa & 4.1 & 19.0 & 0.06  & \citet{2001ApJ...551L.147M}, \citet{2006ApJ...648L.135M} \\
		GRS~1124-684 & GU Mus & 10.4 & 20.5 & 0.9 & \citet{1992ApJ...399L.145R}, \citet{2001AJ....122..971G} \\
		4U~1543-475 & IL Lupi & 26.8 & 16.6 & 1.5 & \citet{1998ApJ...499..375O} \\
		XTE~J1550-564 & V381 Nor & 37.0 & 22.0 & 4.75 & \citet{2002ApJ...568..845O} \\
		GRO~J1655-40 & V1033 Sco & 62.9 & 17.3 & 4.03 & \citet{1995Natur.374..701B}, \citet{1997ApJ...477..876O} \\
		GX~339-4 & V821 Ara & 42.1 & 19.2 & 3.4 & \citet{2000MNRAS.312L..49K}, \citet{2003ApJ...583L..95H} \\
		H~1705-250 & V22107 Oph & 12.5 & 21.5 & 1.2 &\citet{2012MNRAS.427.2876Y}, \citet{1996ApJ...459..226R} \\
		SAX~J1819.3-2525 & V4641 Sgr & 67.6 &13.7 & 0.9 &\citet{2001ApJ...555..489O} \\
		XTE~J1859+226 & V406 Vul & 9.2 & 23.29 & 1.8 &\citet{2002MNRAS.334..999Z} \\
		GRS~1915+105 & V1487 Aql & 804.0 & - \tnote{b} & 19.6 & \citet{2001Natur.414..522G}, \citet{Corbel} \\
		GS~2000+251 & QZ Vul & 8.3 & 21.7 & 4.0 & \citet{1995MNRAS.277L..45C}, \citet{1991MNRAS.249..573C} \\
		GS~2023+338 & V404 Cyg & 155.3 & 18.5 & 3.3 & \citet{1994MNRAS.271L..10S}, \citet{1995MNRAS.277L..45C} \\
        XTE~J1650-500 & - & 7.6 & 24.0 & 4.65 & \citet{2004ApJ...616..376O}, \citet{2002ATel..104....1G}, \citet{2001IAUC.7710....2A}\\
         GS~1354-64 & BW Cir & 61.1 & 21.5 & 3.1 & \citet{2009ApJS..181..238C}, \citet{1990ApJ...361..590K}  \\       
		\hline
\end{tabular}
    \begin{tablenotes}
    	\item[a] Magnitude in quiescence
    	\item[b] For calculating the detection fractions, the absolute magnitude of GRS~1915+105 was assumed to be $M_V$ = 0
    \end{tablenotes}
\end{threeparttable}
\end{table*}

The modelling of the implied period distribution of the sources takes into account the probability of detecting the companion in quiescence, the probability of detecting an outburst with an all-sky monitor based on its location in the Galaxy, as well as the probability of the source having a detectable outburst based on the recurrence time and duration of the outburst. It is also worth mentioning that few of the BHXBs have repeated outbursts, and their recurrence times remain a key unknown. For more on the outburst recurrence times, see Section \ref{sec:prob}. \\

\subsection{Extinction of Optical Counterpart}
\label{sec:optical}

To determine the fraction of binaries for which the optical counterparts are too dim to obtain radial velocity mass measurements due to the effects of extinction from the interstellar medium, the three-dimensional extinction model of \citet{2006A&amp;A...453..635M} was used. The catalogue contains extinction profiles along over 64000 lines of sight in the regions of $\vert l \vert \leq 100^{\circ}$ and $\vert b \vert \leq 10^{\circ}$ using 2MASS data.  Linear interpolation of the data from the catalogue was used to determine the amount of extinction to each of the 25000 locations determined as described in Section \ref{sec:galmod}. \\

\begin{figure}
	\includegraphics[width=\columnwidth]{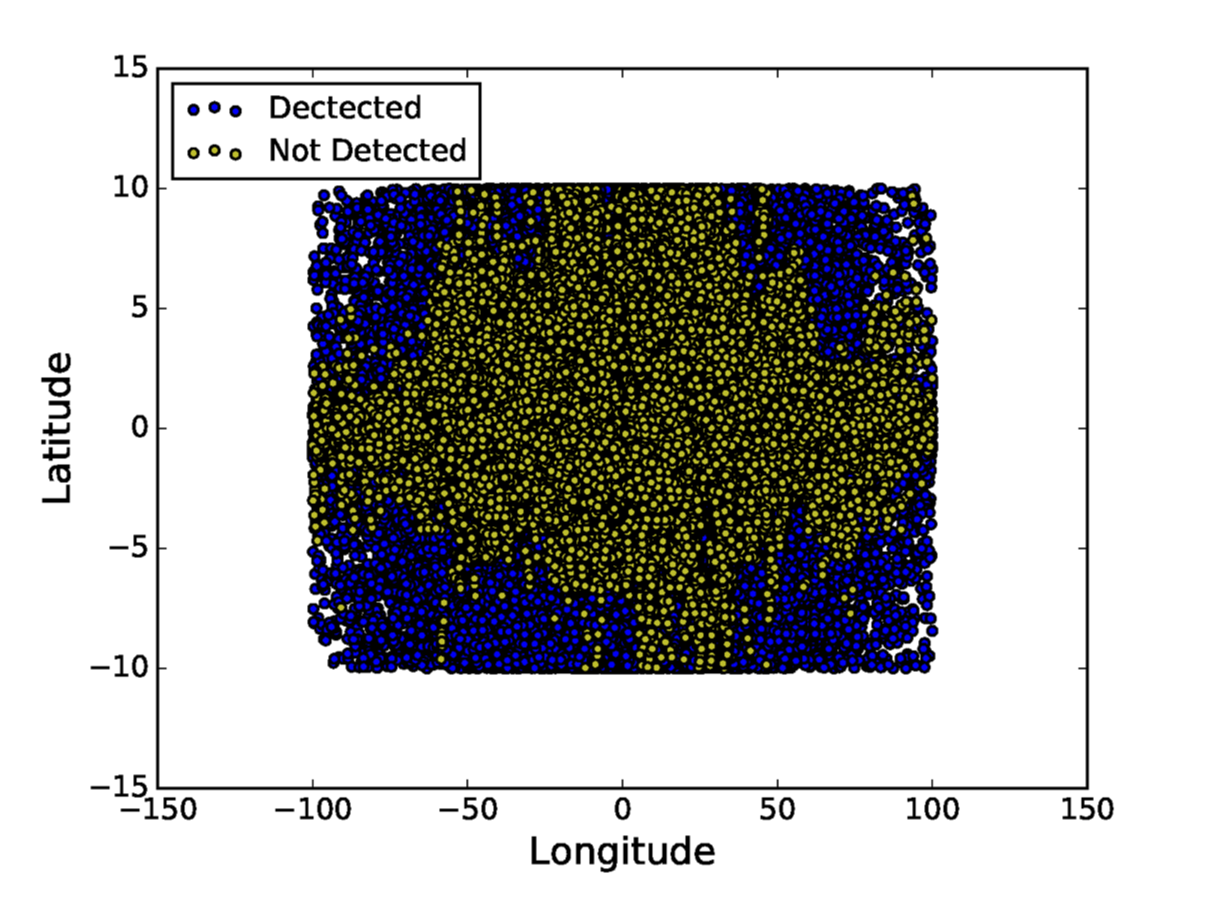}
    \caption{An example plot showing the detection of the optical counterpart for the source XTE J1150-564. The locations indicated by the blue filled circles are the locations where the optical counterpart can be observed. The yellow filled circles indicate locations where the optical counterpart cannot be detected due to high levels of extinction along that line of sight.}
    \label{fig:detect_eg}
\end{figure}

For the optical counterpart, a cut off magnitude of V=22 was selected. After accounting for extinction, sources along each line of sight brighter than this magnitude are considered to be bright enough for radial velocity measurements. The quiescent magnitudes used for the sources are listed in Table~\ref{tab:bh_prop}. Fig. \ref{fig:detect_eg} shows an example of the process of calculating the fraction of optical counterparts that are visible. The fraction of companion stars that are visible in the optical for each source is listed in Table~\ref{tab:det_frac}. \\

\subsection{Detection of the X-ray outburst}
\label{sec:xrays}

To determine if the outburst would be visible, the recorded values of the peak fluxes seen from the sources was used to calculate the peak luminosities of the X-ray outburst. The outburst was considered to be  detectable if the flux from the source exceeded 2.3 counts per second (corresponding to a 30 mCrab flux for RXTE ASM) after accounting for the absorption column along that line of sight. \\

The conversion from X-ray flux counts per second was done using the NASA HEASARC tool PIMMS. The values of hydrogen column density were obtained from \citet{2006A&amp;A...453..635M}, using N$_H$ = 1.79 $\times$ 10$^{21}$A$_V$cm$^{-2}$ mag$^{-1}$ \citep{1995A&amp;A...293..889P}. A standard photon index of $\Gamma$ = 1.7 was assumed for all sources. The fraction of X-ray outbursts that can be detected for each source is listed in Table \ref{tab:det_frac}. 

\subsection{Probability of outburst detection}
\label{sec:prob}
The probability of a source having an outburst that is detectable by an all-sky monitor depends on both the recurrence time for outbursts and the duration of the outburst. The probability of a source having at least one outburst that could have been detected is given by: 

\begin{equation}
\mathcal{P} (\geq 1 {\rm outburst}) = 
\begin{cases}
	 t_{survey} / t_{rec} & \text{if } t_{rec} > t_{survey}\\
	 1			& \text{otherwise}
\end{cases}
\label{eq:trec}
\end{equation}

where $t_{rec}$ is the outburst recurrence time and $t_{survey}$ is the survey length. The survey length was taken to be 15 years, the lifetime of the Rossi X-Ray Timing Explorer All Sky Monitor (RXTE ASM). The outburst recurrence time $t_{rec}$ was assumed to be:
\begin{equation}
t_{rec} =
\begin{cases}
	 M_D/|\dot{M_2}|  & \text{if } L < 0.02 L_{Edd}\\
	 N \times M_D/|\dot{M_2}| 		& \text{otherwise}
\end{cases}
\end{equation}

 where $M_D$ is the disc mass, $\dot{M_2}$ is the rate at which mass is transferred from the companion and N is a factor that accounts for an increase in the recurrence rate due to mass lost in the accretion disc due to wind. In this paper, we ran the simulations using N=2, N=3 and N=10, which corresponds to a mass loss of 50\%, 66\% and 90\% respectively. If a source has an outburst during a survey, for it to be observed the outburst must be visible for at least 1 day \citep{2014MNRAS.437.3087K}. For outbursts lasting less than 1 day, the probability of being observed decreases. The probability of observing an outburst $\mathcal{P}_{obs}$ is given as:

\begin{equation}
\mathcal{P}_{obs} = 
\begin{cases}
	 t_{det} / 1 day & \text{if } t_{det} < \text{1 day}\\
	 1			& \text{otherwise}
\end{cases}
\end{equation}

where $t_{det}$ is the observable outburst duration. For a limiting flux $F_{lim}$, this is determined to be the time at which the luminosity falls below $L_{lim} = 4\pi d^2 F_{lim}$. We have defined $F_{lim}$ to be 10mCrab, corresponding to the daily exposure sensitivity of RXTE. The luminosity is taken to be L = $\eta \dot{M} c^2 $ where $\eta$ is the radiative efficiency and $\dot{M}$ is the accretion rate onto the central compact object.  The efficiency at high luminosities can be approximated to be $\eta$ = 0.1. However, at luminosities below a few percent of the Eddington luminosity, the accretion becomes radiatively inefficient. Thus we assume a transition to lower efficiencies used in \citet{2014MNRAS.437.3087K} as: 

\begin{equation}
\label{eq:efficiency}
\eta = 0.1 \left(\frac{\dot{M}}{f\dot{M}_{Edd}}\right)^n
\end{equation}

for $L \leq fL_{Edd}$ where $f = 0.02$ and $n = 1$. This transition to lower efficiencies can reduce the duration for which the outburst is visible. In the cases of binaries with short orbital periods, this can render the outburst undetectable in all but the most favourable of locations in the Galaxy. \\

It has been shown by \citet{1998MNRAS.293L..42K} that the central accretion rate can be described by an initial exponential fall off followed by a linear decay:

\begin{equation}
\dot{M} = \left(\frac{3\nu}{B_{n}}\right)^{1/2} \left[ M_h^{1/2}(T) -\left(\frac{3\nu}{B_{n}}\right)^{1/2} (t-T) \right]
\end{equation}

where $M_h$ is the irradiated mass at a given time t and the constant $B_{n}$ was set to $10^5$ \citep{2014MNRAS.437.3087K}, 
\begin{equation}
M_h(t) = \frac{\rho R_d^3}{3} {\rm exp} \frac{-3\nu t}{R_d^2}
\end{equation}

and T is the time taken for the the irradiated radius to drop below the disc radius (T=0 for long period systems)
\begin{equation}
T = \frac{R_d^2}{3\nu} {\rm log} \frac{B_n\nu \rho}{R_d}
\end{equation}

The value of density $\rho = 10^{-8} {\rm g cm^{-3}}$  was chosen following \citet{1998MNRAS.293L..42K}, where the density was shown to be independent of the radius, making it suitable for the wide range of orbital periods used. \\

The viscosity is taken to be $\nu = \alpha_{h} c_{s} H = \alpha_{h} c_{s} (H/R) R_{d}$ \citep{1973A&amp;A....24..337S, 1981ARA&amp;A..19..137P} where $c_{s}$ is the speed of sound and H is the scale height of the disc. We have chosen a value of $\nu = 2.2 \times 10^{14}{\rm cm^2 s^{-1}}$, as this produces a reduction in the radiative efficiency of BHXBs with an orbital period less than 5 hours. \\

In addition to the above, it must be noted that all sky monitors cannot observe 100\% of the sky at any given time due to the location of the Sun. It is thus possible for binaries that are close to the ecliptic and exhibit short period outbursts to be obscured by the Sun. The probability of detecting a binary having an outburst that is partially obscured is given by:

\begin{equation}
\mathcal{P}_{Sun} =
\begin{cases}
	1.0 - \frac{t_{obsc} - t_{det} + 1}{365.25} & \text{if } t_{det} < t_{obsc}\\
	 1.0			& \text{otherwise}
\end{cases}
\end{equation}

 where $t_{obsc}$ is the number of days per year {\bf that} the binary is behind the Sun.

Thus the total probability of having an outburst and observing it is given by:
\begin{equation}
\label{eq:det}
\mathcal{P}_{det} = \mathcal{P}_{obs} \times \mathcal{P} (\geq1 {\rm outburst}) \times \mathcal{P}_{Sun}
\end{equation}
The probability of detecting an outburst from each of the 17 BHXBs is listed in Table \ref{tab:det_frac}.

\begin{table}
	\centering	
       \caption{Detection fractions of BHXBs.}
    \label{tab:det_frac}
	\begin{tabular}{lccc} 
		\hline
		Source Name & \thead{Detection \\ Fraction \\ (X-Rays)} & \thead{Detection \\ Fraction \\ (Optical)} & \thead{Outburst \\ Detection \\ Probability} \\
		\hline
		V518 Per & 0.0025 & 0.99 & 1.0 \\
		V616 Mon & 0.025 & 1.0 & 0.38\\
		MM Vel & 0.0051 & 0.99 & 1.0\\ 
		KV UMa & 0.0085 & 0.0034 & 1.0 \\
		GU Mus & 0.0078 & 1.0 & 0.38\\
		IL Lupi & 0.38 & 1.0 & 0.27\\
		V381 Nor & 0.22 & 1.0& 0.21 \\
		V1033 Sco & 0.38 & 1.0 & 0.27 \\
		V821 Ara & 0.33 & 1.0 & 0.37 \\
		V2107 Oph & 0.051 & 1.0 & 0.69 \\
		V4641 Sgr & 0.44 & 1.0 & 0.13 \\
		V406 Vul & 0.011 & 1.0 & 0.56 \\
		V1487 Aql & 0.40 & 1.0 & 0.053 \\
		QZ Vul & 0.072 & 1.0 & 0.43 \\
		V404 Cyg & 0.34 & 1.0 & 0.11 \\	
        J1650-500 & 0.10 & 0.75 & 0.55 \\
        BW Cir & 0.35 & 1.0 & 0.25 \\
		\hline
	\end{tabular}
\end{table}

\begin{table}
	\centering
    \caption{Power law fits to number of detected sources vs. $P_{orb}$}
    \label{tab:fitvals}
    \begin{threeparttable}
    \begin{tabular}{lcr}
    \hline
    Distribution & No. of sources & Power Law Index  \\
    \hline
    Data & 7 & 2.45 $\pm$ 1.3 \\
    Spike \tnote{a} & 500 & 1.60 $\pm$ 0.09 \\
    Flat \tnote{b} & 500 & 2.38 $\pm$ 0.10 \\
    Spike &  275 & 1.62 $\pm$ 0.12 \\
    Flat & 275 & 2.38 $\pm$ 0.13\\
    Spike & 50 & 1.61 $\pm$ 0.31\\
    Flat & 50 & 2.36 $\pm$ 0.32\\
    \hline   
    \end{tabular}
        \begin{tablenotes}
    	\item[a] Drawn from an initial orbital period distribution where all the binaries start at a 20 hour period.
    	\item[b] Drawn from an initial orbital period distribution that was logarithmically flat between 2 and 20 hours
    \end{tablenotes}
    \end{threeparttable}
\end{table}

\subsection{Implied Orbital Period Distribution using Observational Data}
\label{sec:implied_period_dist}

\begin{figure*}
	\includegraphics[width=\textwidth]{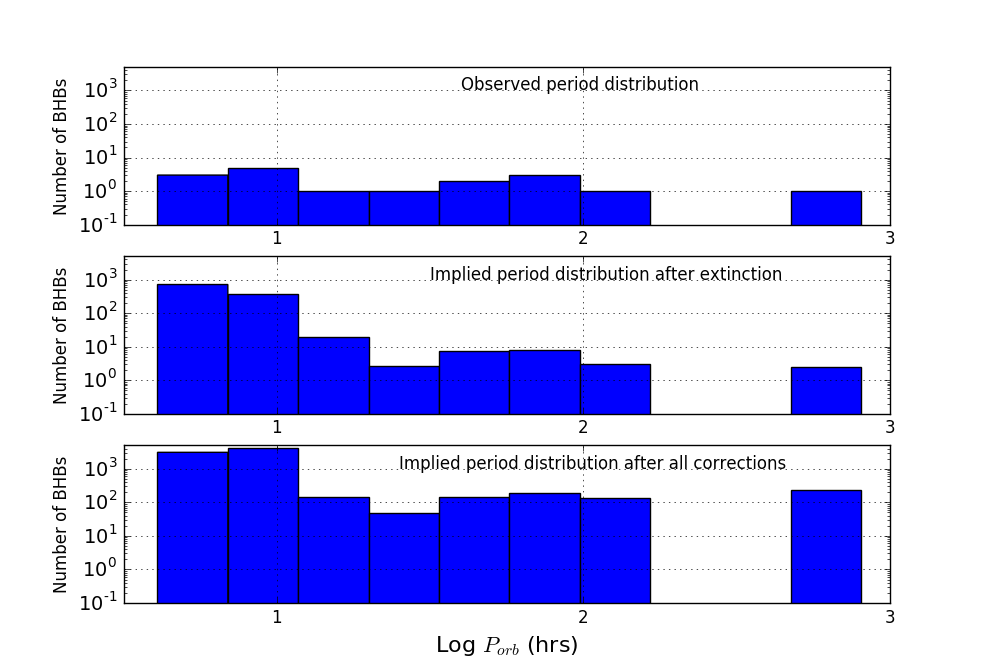}
    \caption{Top: Histogram of the observed period distribution of BHXBs. Middle: Distribution after accounting for extinction of optical counterpart. Bottom: Distribution after accounting for extinction of optical counterpart, absorption of X-ray flux and the probability of detecting the X-ray outburst.}
    \label{fig:Impl_dist}
\end{figure*}

Using the detection fractions shown in Table \ref{tab:det_frac}, the implied orbital period distribution was plotted. The results are shown in Fig. \ref{fig:Impl_dist}.\\ 

To compare the results with the distribution expected from theoretical models of magnetic braking, we fit a power law slope to the results obtained after correcting for the selection effects detailed above. While the fitting of a constant cannot be excluded in a statistically significant manner, it is expected that the distribution follows a power law due to the physical processes dominating the evolution of the system (see Equation \ref{eq:magbr}). For this fitting, any sources with an orbital period greater than 10 hours were excluded, since the evolution of these sources is likely to be dominated by expansion of the donor star, rather than by magnetic braking. The bin sizes used are unequal to ensure that none of the bins had a count of zero, and the error bars shown are Poissonian. \\

\begin{figure}
	\includegraphics[width=\columnwidth]{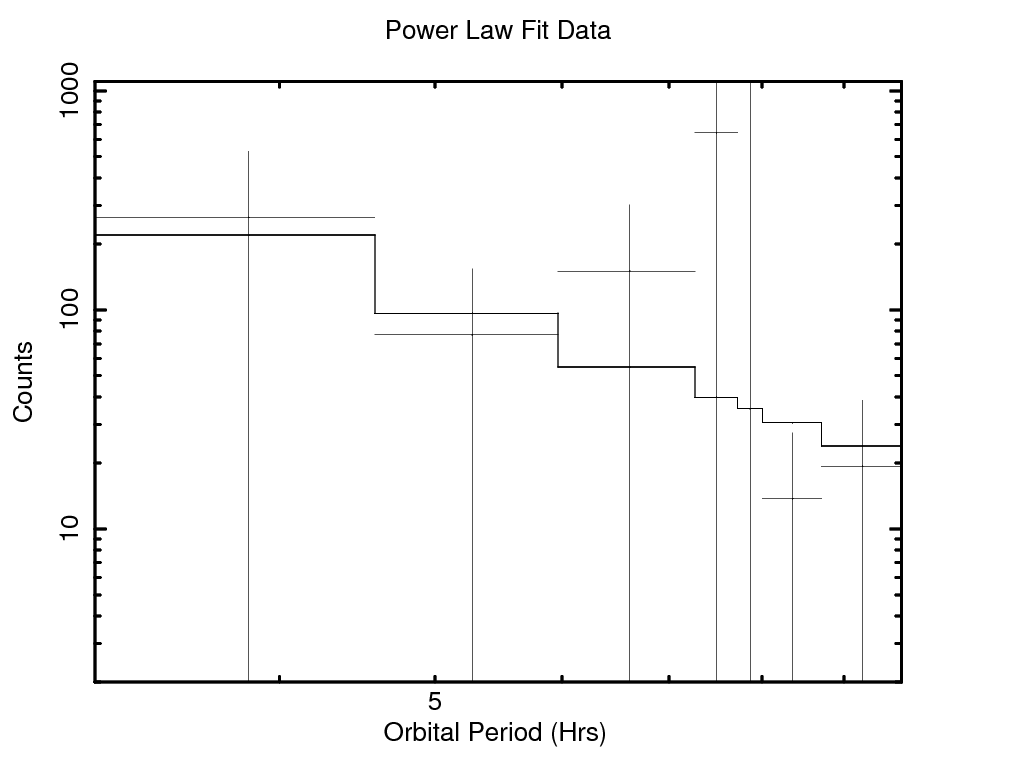}
    \caption{A power law model fit to the predicted number of black hole binaries in the Galaxy with an orbital period of less than 10 hours. }
    \label{fig:plfit_sources}
\end{figure}

The fitting of the power law was done using XSPEC \citep{1996ASPC..101...17A}. For the fitting, data from the observed BHXBs were converted to a spectrum file format using the flx2xsp routine in FTOOLS. Combining the optical detection fractions, the X-ray detection fractions and the probability of detecting the initial outburst gives the final fraction of the total BHXB population that can be observed. This final fraction was used as the corresponding response matrix for use with XSPEC. The results of the fit are shown in Fig. \ref{fig:plfit_sources}. We obtained a best fit value of $2.45 \pm 1.3$ for the plot  assuming a 50\% mass loss due to disc winds (i.e. N=2 in Equation \ref{eq:trec}). It was found that changing the amount of mass lost did not significantly alter the best fit value of the slope. It does, however, affect the normalisation of the power law. These results are further discussed in Section \ref{sec:discussion}. The large error bars on the fit are due to the small number of binaries with an orbital period less than 10 hours.

\section{Monte Carlo Simulation}
\subsection{Generation of simulated sources}

We generated a population of 250,000 simulated sources by drawing from two simplified initial period distributions. The first is the spike distribution is where all the sources start at an orbital period of 20 hrs and the second is the flat distribution where the orbital periods are logarithmically distributed between 2 hrs and 20 hrs. The orbital period of these objects decay as a result of mass transfer driven magnetic braking and gravitational radiation that can be described using:

\begin{equation}
\label{eq:magbr}
-\dot{M_2} =
\begin{cases}
	10^{-10} \left( \frac{P_{orb}}{2} \right)^{-2/3} M_{\odot} yr^{-1} & \text{if P(hr) < 2}\\
	6 \times 10^{-10} \left( \frac{P_{orb}}{3} \right)^{5/3} M_{\odot} yr^{-1} & \text{if P(hr) > 3}
\end{cases}
\end{equation}

 as shown by \citet{1988QJRAS..29....1K}. In the case of the orbital period being greater than 3 hours, the mass transfer is dominated by magnetic braking. For periods less than 2 hours, this transfer is driven by gravitational radiation. For periods between 2 \& 3 hours, there is likely to be little to no mass transfer as the star's structure changes as it becomes fully convective, and it briefly stops filling its Roche lobe.\\

\begin{figure}
	\includegraphics[width=\columnwidth]{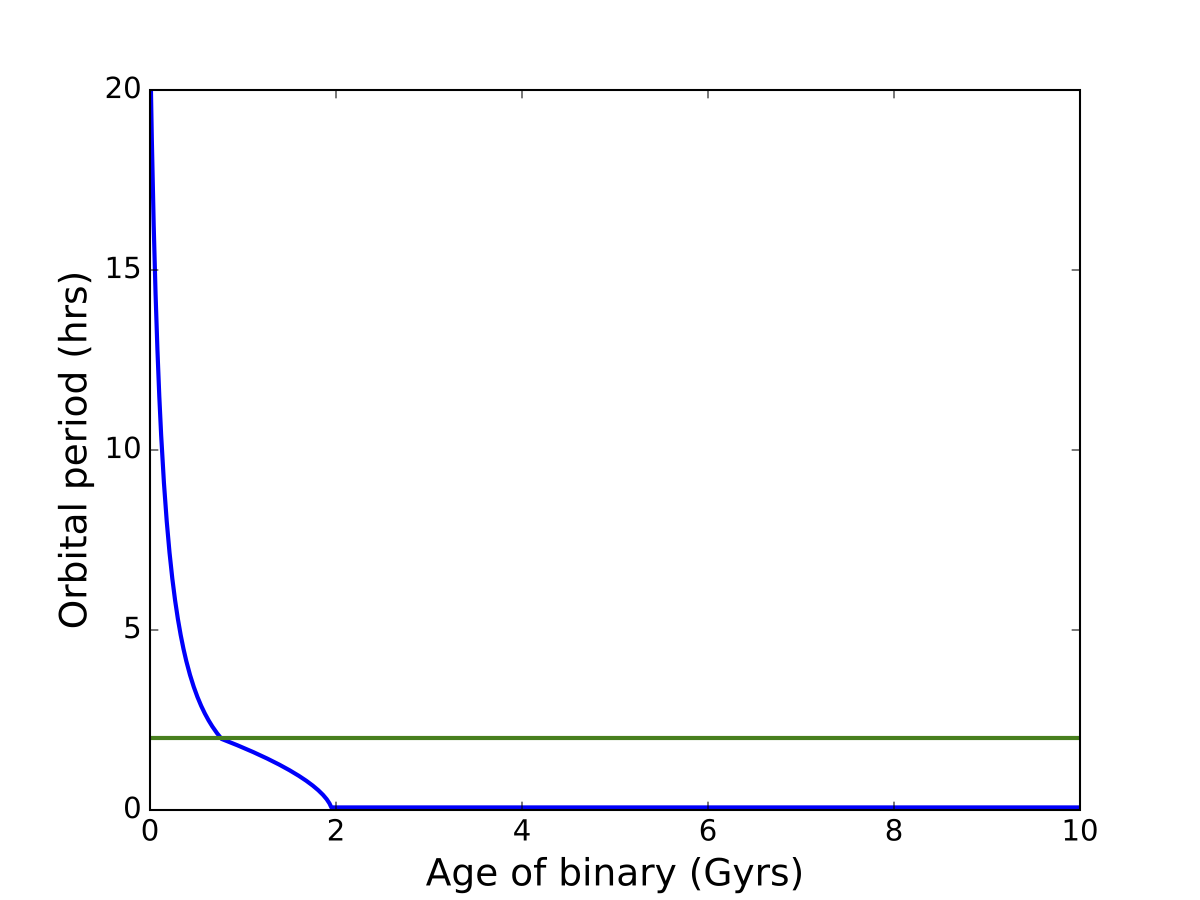}
    \caption{Plot of orbital period decay via magnetic braking and gravitational radiation for a system starting at an orbital period of 20 hours. The green horizontal line indicates a 2 hour orbital period, below which it is harder to observe the sources unless they are at a close distance.}
    \label{fig:period_decay}
\end{figure}

The age of each binary was selected randomly from a uniform distribution between 0 and 10 Gyrs, and its current orbital period was determined using  Equation \ref{eq:magbr}. A plot of the current orbital period of the binary as a function of its age is shown in Fig. ~\ref{fig:period_decay}. In these simulations, all the binaries were assumed to have a primary of mass equal to 8$M_{\odot}$. The companion mass was taken to be $m_2 = 0.1P_{orb}$(hrs) for $P_{orb}$ < 10 hours, and $m_2 = 0.7$ for $P_{orb}$ > 10 hours. This way, the current orbital period distributions of the two samples were generated. \\

\subsection{Optical counterpart detection }
The luminosity of the companion star was calculated assuming that the star is a lower main sequence star that just fills its Roche lobe, with a temperature of 3500K. This corresponds to the temperature of typical KV type star. The Roche lobe radius of the secondary star is taken to be:

\begin{equation}
\frac{R_{L2}}{a} = \frac{0.49q^{2/3}}{0.6q^{2/3}+ \text{ln}(1+q^{1/3})}
\end{equation}
\citet{1983ApJ...268..368E}, where the binary separation {\bf $a$} is given by
\begin{equation}
a = 3.5 \times 10^{10}m_1^{1/3}(1+q)^{1/3} P_{orb}^{2/3} \text{cm}
\end{equation}

A 50\% contribution of light from the accretion disc in quiescence was assumed and the locations of the binaries were selected randomly (weighted by stellar density). The effect of extinction was calculated as described in Section \ref{sec:optical}. 

\subsection{Probability of an X-ray outburst and its detection }
The probability of detecting the X-ray outburst for these generated sources was calculated using the method detailed in Section \ref{sec:xrays}. Once the detection fractions were calculated, a sample of 500 generated sources were randomly selected from each of the two distributions. \\

The peak luminosity of the X-ray outburst was taken to be $L_{peak} = \eta c^2 \rho \nu R_d$, where  $R_d$ is the disc radius and $\eta$ is the radiative efficiency given by:
\begin{equation}
\label{eq:efficiency}
\eta = 0.1 \left(\frac{\dot{M}}{f\dot{M}_{Edd}}\right)^n
\end{equation}
for $L \leq fL_{Edd}$ where $f = 0.02$ and $n = 1$ The values of $\nu$ and $\rho$ used are the same as described in Section \ref{sec:prob} . 

The disc radius $R_d$ was approximated as $R_d \approx 0.7R_{L1}$ where $R_{L1}$ is the Roche radius of the primary as given by: 

\begin{equation}
\frac{R_{L1}}{a} = \frac{0.49q^{-2/3}}{0.6q^{-2/3}+ \text{ln}(1+q^{-1/3})}
\end{equation}

\subsection{Implied Orbital Period Distribution using Simulated Data}

\begin{figure*}
  \centering
  \subfloat[Flat Distribution.]{\includegraphics[width=\columnwidth]{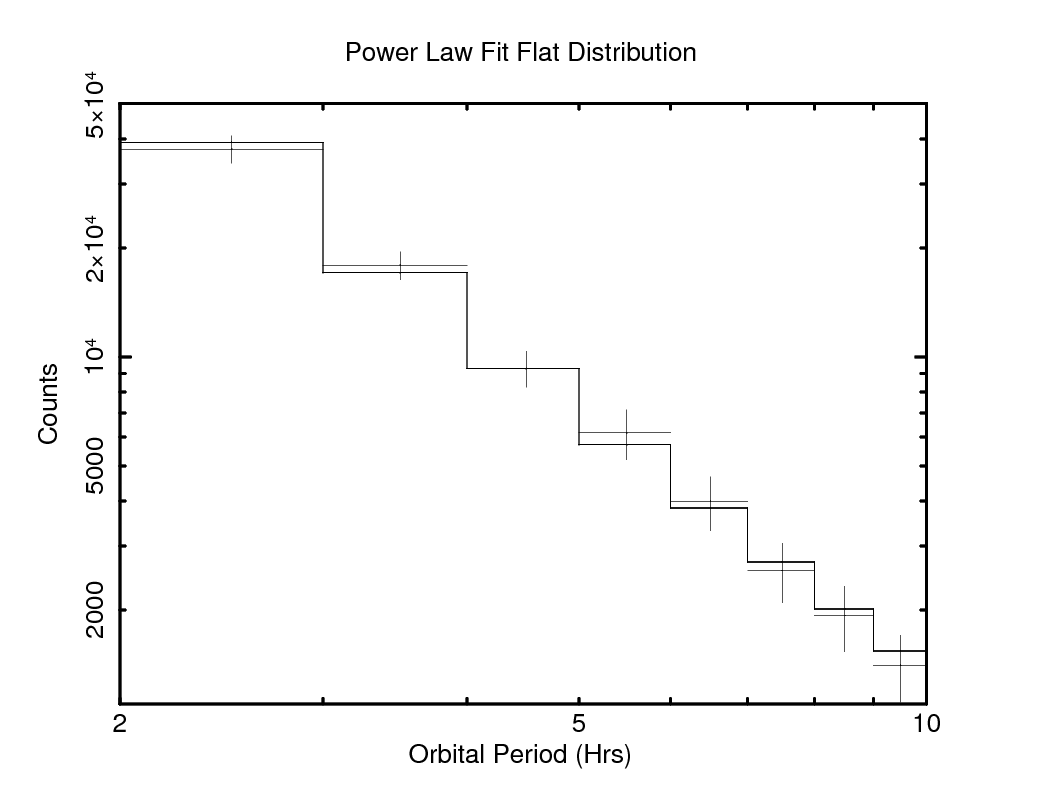}\label{fig:flat500}}
  \subfloat[Spike Distribution.]{\includegraphics[width=\columnwidth]{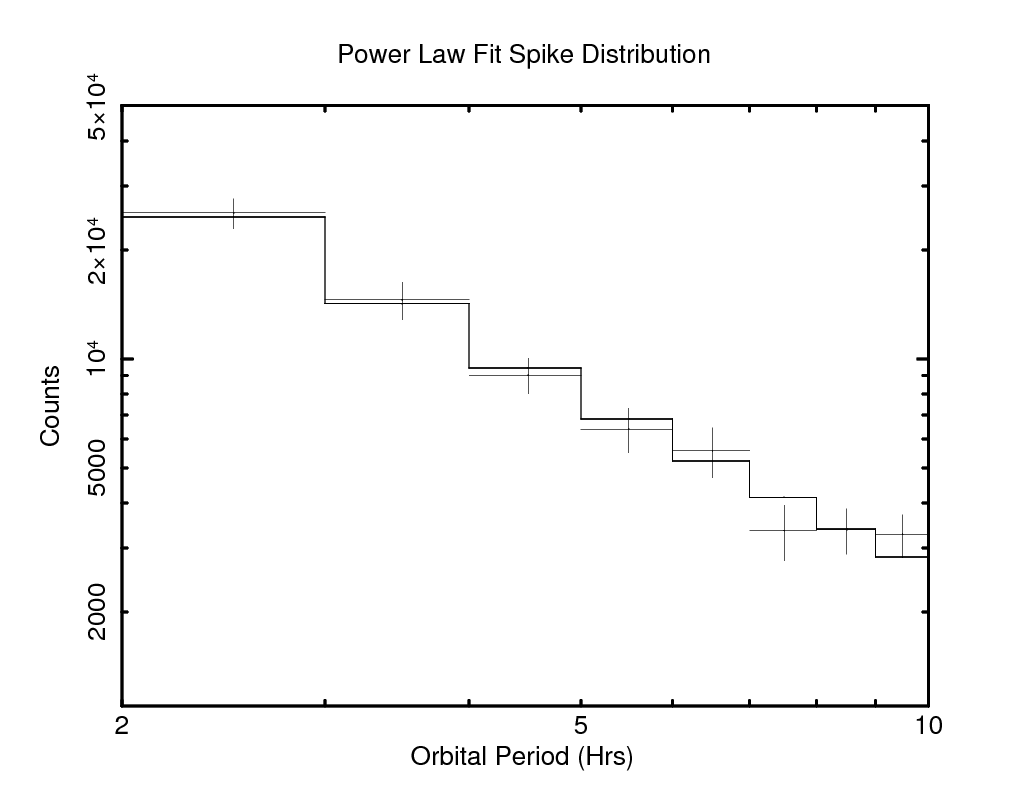}\label{fig:spike500}}
  \caption{The best power law fit obtained using XSPEC for the orbital period distribution of 500 randomly generated black hole binaries with periods between 2 and 10 hours. The binaries in the left panel were drawn from an initial orbital period distribution that was logarithmically flat between 2 and 20 hours and decay via magnetic braking. The binaries in the right panel were drawn from an initial orbital period distribution where all the binaries start at a 20 hour period and decay via magnetic braking.}
  \label{fig:bestfit500}
\end{figure*}

The best fits for the power law index for the plots of number of detected sources vs. the orbital period using the magnetic braking rate as shown in Equation(\ref{eq:magbr}) are shown in Table \ref{tab:fitvals}. The plots of the best fits using a sample of 500 detected sources from each distribution can be seen in Fig.\ref{fig:bestfit500}. 

\subsection{Effect of mass transfer on the companion}

The fit to the slope of the orbital period distribution plotted against the number of sources depends on the shape of the distribution that the sample is drawn from, as well as the relationship between the mass transfer rate and the orbital period of the system. In addition, due to the mass transfer that occurs during the lifetime of the donor star, the star can be expected to have a radius greater than the radius of an isolated main sequence star of the same mass \citep{1996ApJ...464L.127K}. This bloating factor can also affect the slope of the plot, making it steeper. \\

The model slope ${dN}/{dP}$ is determined by the amount of time a binary spends at a particular orbital period $\dot{P}^{-1}$ (i.e. the inverse of the rate of change of the orbital period). From the mass-period relationship, it can be assumed that: 
\begin{equation}
\frac{\dot{M_2}}{M_2} \propto \frac{\dot{P}}{P}
\end{equation}
Since $P \propto M_2$ \citep{1988QJRAS..29....1K}, this gives us: 
\begin{equation}
\frac{dN}{dP} \propto \dot{M_2}^{-1}
\end{equation}

Since Equation(\ref{eq:magbr}) does not take into account of the effect of mass transfer on the donor star, using Equation 9 from \citep{1996ApJ...464L.127K} results in:

\begin{equation}
\frac{dN}{dP} \propto \left( \frac{P_{orb}}{3} \right)^{-5/3} \hat{m}_2 ^{-7/3}
\end{equation}

where $ \hat{m}_2$ = $M_2$/$M_2$(MS). $M_2$ is the mass of the companion star, and $M_2$(MS) is the mass of a main-sequence star that would just fill the Roche lobe at the current period.

\section{Discussion}
\label{sec:discussion}
By fitting the results of the implied period distribution from the observed binaries, a best fit value of $2.45 \pm 1.3$ was obtained for the plot assuming a 50\% mass loss due to disc winds (i.e. N=2 in Equation \ref{eq:trec}). Changing the amount of mass lost to 66\% and 90\% did not significantly alter the best fit value of the slope.  \\

From the results in Table \ref{tab:fitvals}, it would appear that the sample drawn from the flat orbital period distribution matches the data. The addition of the effect from the bloating of the companion star steepens the value of the slope by approximately 0.9, making the best fit value 3.28. However, due to the large error on the best fit to the slope of the observed data, the fit of the  sample drawn from the spiked orbital period distribution cannot be excluded.\\

By integrating the implied period distribution to estimate the number of short period black hole binaries, we predict that approximately 600 black hole binaries are present in the Galaxy with an orbital period between 3 and 10 hours. Altering the amount of mass lost in winds to 66\% yields an estimate of $\sim$1100 binaries, and a loss of 90\% of the mass yields an estimate of $\sim$3700 binaries. It was estimated by \cite{2011ApJ...743...26F} that 90\% of the mass was lost due to disc wind and/or jets. This finding is based on estimating the mass transfer rate of the binary from the ultraviolet luminosity of the system under the assumption that the UV comes from the accretion stream impact spot on the outer disc in A0620-00, and then comparing with the duty cycle of the outburst, and is also supported by measurements of strong disc winds in the outbursts of X-ray binaries (e.g. \citealt{2009arXiv0903.4173N}). It has been alternatively suggested that the UV emission may come from closer to the black hole in this system, in which case the inferred mass transfer rate would be much lower, and such strong disc winds would not be necessary (\citealt{2012ApJ...749....3H}), although such a result would not explain the actual observations of strong disc winds in outburst.  Thus it is clear that a deeper understanding of the outburst duty cycle is needed to accurately estimate the total number of black hole binaries in the Galaxy. \\

By extrapolating the fit down to 2 hours, we predict that there are $\sim 200-3000$ binaries with periods between 2 and 3 hours. The range of the values are determined using the upper and lower limits of the powerlaw index estimate as shown in Table \ref{tab:fitvals}. Our estimate for the total number of BHXBs in the Galaxy is consistent with theoretical estimates of numbers between 10$^2$ and 10$^4$ obtained using population synthesis codes (\citealt{1992ApJ...399..621R}, \citealt{1994ASPC...56..196R}, \citealt{1997A&amp;A...321..207P}, \citealt{2006A&amp;A...454..559Y}). However, \citet{2016ApJ...825...10T} have detected a radio source in the direction of M15 that they have suggested is a BHXB in quiescence. If correct, it would imply a much larger population of quiescent BHXBs (2.6 $\times$ 10$^4$ - 1.7 $\times$ 10$^8$). While this could suggest a new channel of black hole formation, it does not fit well with our predictions. It must also be emphasized that X-rays have not been detected from this source, and thus the estimate of a larger population must be treated with caution. While this discovery could be anomalous, it has been suggested  that binaries lack outbursts at accretion rates lower than those predicted by magnetic braking. This could be the case for short period binaries that exist in the period gap which, as discussed by \citet{2013MNRAS.428.1335M} may have low but non zero accretion rates, as well as for the very long period systems suggested by \citet{1999ApJ...513..811M}. Thus it is possible that there exists a hidden population of persistently quiescent BHXBs. \\

To estimate the number of binaries that would have to be detected and have dynamical mass estimates in order to determine which of the two distributions is closest to the true orbital period distribution, we ran the Monte Carlo simulation detailed in the previous section, drawing a different number of detectable sources each time. With a sample of 50 detected binaries, it is possible to distinguish between the two sample orbital period distributions at the 1-sigma level. A sample of approximately 275 binaries will be required to distinguish the two at the 3-sigma level. The next generation of X-ray missions such as LOFT (\citealt{2012ExA....34..415F}) and STROBE-X (\citealt{2017arXiv170903494W}) are thus extremely important for identifying more of these short period systems and understanding their underlying distribution.\\

The model used to predict the peak luminosities of the X-ray outburst produces a decrease in radiative efficiency at short orbital periods (P$_{orb} \leqslant$ 5 hours). However, the predicted luminosity for XTE J1118+480 is higher than the luminosity derived from the observed peak flux from the system. This was also noticed by \citet{2010ApJ...718..620W} who noted that this could be due to advection. If this is true in the case of all BHXBs with periods less than 5 hours, then the detection of the X-ray outburst becomes the limiting factor, as opposed to the detection of the optical counterpart. In this case, the availability of sensitive all-sky monitors is the most effective tool in searching for this population of very faint, short period BHXBs. This can already be seen by the large number of short period black hole candidates being detected by new, more sensitive Wide Field Monitors. However, a number of these candidates cannot be confirmed due to the faintness of the companion in quiescence and an inability to obtain detailed spectroscopy for these objects. A possible solution to the problem in obtaining dynamical confirmation for short period objects could be the relation proposed by \citet{2016ApJ...822...99C} that allows the determination of the mass of the binary using H$_\alpha$ emission lines, rather than absorption lines. Using this method could extend the search for short period BHXBs to fainter limits, allowing us to confirm a larger fraction of candidate systems. \\

It is also worth noting that the addition of natal kicks to the simulation will result in an increase in the scale height of the BHXBs. See e.g. \cite{2004MNRAS.354..355J}, \citet{2015MNRAS.453.3341R} for a discussion of the scale height distribution of BHXBs and corresponding evidence that most form with large natal kicks. This is likely to result in a larger fraction of binaries being detected due to the lower levels of extinction away from the Galactic plane. This can be seen by the detection of short period systems at high galactic latitudes by MAXI (\citealt{2012PASJ...64...32Y}, \citealt{2013Kuulkers}) such as MAXI J1659-152 (\citealt{2010ATel.2873....1N}, \citealt{2013Kuulkers}), MAXI J1836-194 (\citealt{2011ATel.3611....1N}, \citealt{2014MNRAS.439.1381R}), MAXI J1305-704 (\citealt{2012ATel.4024....1S}, \citealt{2017symm.conf...45S}) and MAXI J1910-057 (\citealt{2012ATel.4140....1U}, \citealt{2014efxu.conf..260Y})

\section{Conclusions}
In this paper we have derived the implied period distribution of low mass BHXBs after correcting for the effects of extinction of the optical counterpart, absorption of the X-ray outburst and the probability of detecting a source in outburst. We have compared the power law fit to the implied orbital period distribution to the distributions produced by drawing a sample from two simple orbital period distribution. Based on the results from the simulation, it is likely that the observed data arise from a flat orbital period distribution. However, due to the small number of observed sources, the possibility of an orbital period distribution arising from the decay of the orbit of a large orbital period (20 hours) cannot be ruled out. With a sample of $\sim$275 detected sources,  it would be possible to differentiate between the two distributions. Our simulation predicts $\sim 200-3000$ binaries with periods between 2 and 3 hours, and an additional $\sim$600 binaries between 3 and 10 hours. This is consistent with numbers predicted using population synthesis models.

\section*{Acknowledgements}

We would like to thank Chris Belczynski for ideas which motivated this paper, and Jorge Casares for useful discussions about finding more quiescent black holes and measuring their masses. We  would also like to thank the anonymous referee for helpful and constructive comments.




\bibliographystyle{mnras}
\bibliography{selection} 







\bsp	
\label{lastpage}
\end{document}